\documentclass[conference]{IEEEtran}
\IEEEoverridecommandlockouts
\usepackage{cite}
\usepackage{amsmath,graphicx,url,times,booktabs,tabularx,amsfonts,siunitx}
\usepackage{algorithmic}
\usepackage{graphicx}
\usepackage{textcomp}
\usepackage{xcolor}
\usepackage{url}
\usepackage{multirow}
\usepackage{siunitx}
\usepackage{microtype}
\usepackage{color}
\DeclareMathOperator*{\argmax}{arg\,max}

\usepackage{gensymb}
\usepackage[defaultcolor=red]{changes}

\def\BibTeX{{\rm B\kern-.05em{\sc i\kern-.025em b}\kern-.08em
    T\kern-.1667em\lower.7ex\hbox{E}\kern-.125emX}}
\begin{document}

\let\OLDthebibliography\thebibliography
\renewcommand\thebibliography[1]{
  \OLDthebibliography{#1}
  \setlength{\parskip}{0pt}
  \setlength{\itemsep}{4.4pt plus 0.1ex}
}

\title{Text-Queried Target Sound Event Localization}

\author{\IEEEauthorblockN{1\textsuperscript{st} Given Name Surname}
\IEEEauthorblockA{\textit{dept. name of organization (of Aff.)} \\
\textit{name of organization (of Aff.)}\\
City, Country \\}}
\author{
      \IEEEauthorblockN{
      Jinzheng Zhao$^{\star}$, Xinyuan Qian$^{\dagger}$, Yong Xu$^{\ddagger}$, Haohe Liu$^{\star}$, Yin Cao$^{\S}$, Davide Berghi$^{\star}$,  Wenwu Wang$^{\star}$
     }
     \\
      \IEEEauthorblockN{$^{\star}$Centre for Vision, Speech and Signal Processing (CVSSP), University of Surrey, UK }
      \IEEEauthorblockN{$^{\dagger}$Department of Computer Science and Technology, University of Science and Technology Beijing, China }
      \IEEEauthorblockN{$^{\ddagger}$Tencent AI Lab, Bellevue, WA, USA }
      \IEEEauthorblockN{$^{\S}$Department of Intelligent Science, Xi’an Jiaotong Liverpool University, China}
      }
\maketitle

\begin{abstract}
Sound event localization and detection (SELD) aims to determine the appearance of sound classes, together with their Direction of Arrival (DOA). However, current SELD systems can only predict the activities of specific classes, for example, 13 classes in DCASE challenges. In this paper, we propose text-queried target sound event localization (SEL), a new paradigm that allows the user to input the text to describe the sound event, and the SEL model can predict the location of the related sound event. The proposed task presents a more user-friendly way for human-computer interaction. We provide a benchmark study for the proposed task and perform experiments on datasets created by simulated room impulse response (RIR) and real RIR to validate the effectiveness of the proposed methods. We hope that our benchmark will inspire the interest and additional research for text-queried sound source localization.
\end{abstract}


\section{Introduction}

Sound event localization and detection (SELD) has attracted increasing interest recently owing to the launch of Detection and Classification of
Acoustic Scenes and Events (DCASE) Task 3, whose objectives are to predict the temporal and spatial activities of sound events jointly. In previous editions, only audio modality (e.g. multi-channel audio in FOA and MIC format) is used. In DCASE 2023, $360 \degree$ videos are added to provide additional information and complementary modality. However, current task settings are constrained to limited class prediction. The number of classes in DCASE 2023 and L3DAS 2023 is 13, covering people speaking, laughter and music, etc, which follow the AudioSet \cite{gemmeke2017audio} ontology. 

Text-prompt based tasks have become prevalent recently as the model works according to the text description and is not constrained to limited classes. In the visual domain, text is used to describe the visual objects for tracking in the image plane in \cite{wang2021towards}. Imagic is proposed in \cite{kawar2023imagic} for text-conditioned image editing. In the audio domain, text queried audio tagging is introduced in \cite{oncescu2021audio}. In \cite{liu2023audioldm}, general sound synthesis is based on the text input. In \cite{liu2022separate, liu2023separate}, general source separation is conducted based on the text description and is not constrained to specific classes. Visual modality is further added in \cite{tan2023language} to assist text-prompted source separation. In \cite{hao2023typing}, text-queried separation is extended to the speech domain. Users can type to describe the speaker they want to separate, such as `the loudest speaker', `the female speaker', etc. In \cite{jiang2023prompt}, text-queried target speech diarization is explored and text is used to choose the target speaker such as `the person who spoke at two seconds', `the male speaker' or `the keynote speaker'. 

\begin{figure}[tbp]
    \centering
    \includegraphics[width=1\columnwidth]{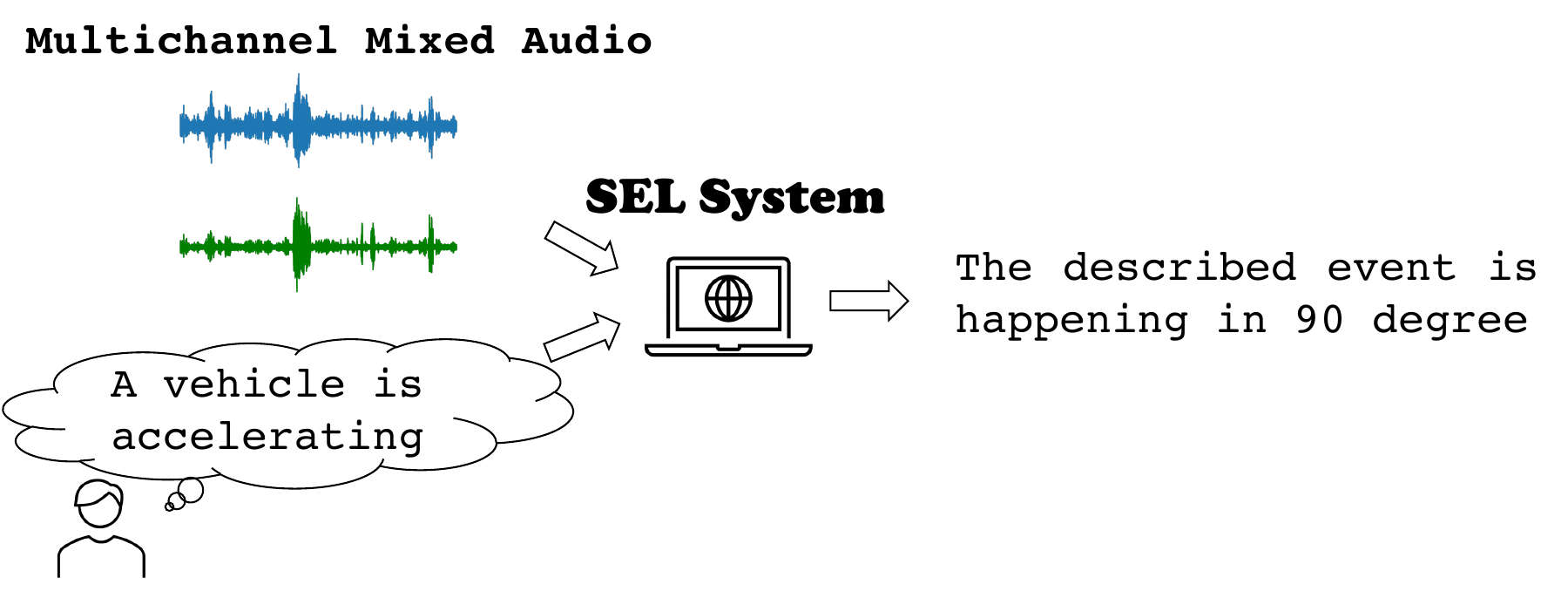}
    \caption{An illustration of the proposed text-queried target sound event localization system. The SEL system takes the input of spatial audio and the user's text description, and predicts the azimuth of the related sound source.}
    \label{sed}
\end{figure}

Inspired by the text-driven tasks, we propose the text-queried target sound event localization, which takes the text description as input and outputs the position of the sound event related to the description, as illustrated in Fig. \ref{sed}. The proposed task requires the model to explore the relationship between the textural information and the spatial information, and locate the target source based on multi-modal correspondence. Some contrastive pretrained models have been proposed to explore the semantic relationships between different modalities.  Contrastive Language–Image Pretraining (CLIP) \cite{radford2021learning} learns the visual-textual semantic information in a self-supervised way. WAV2CLIP \cite{wu2022wav2clip} leverages the pretrained CLIP and distills knowledge from the visual-textual domain into the audio-textual domain. In \cite{wu2023large}, a contrastive language-audio pretraining model (CLAP) is proposed to model the audio-textual representation. Here, we explore how the pretrained multimodal representation models help our proposed task. To our knowledge, this is the first attempt to explore the open set text-queried target sound source localization.

The remainder of the paper is organised as follows. Section \ref{related} recaps the relevant literature such as sound event localization and detection, and multimodal representation learning. Section \ref{define} formulates the problem. Section \ref{m} introduces the proposed multimodal fusion model. Section \ref{implement} provides the training details and Section \ref{experiment} presents and analyzes the experiments. Section \ref{con} points out the limitations and concludes the paper.

\section{Related Work}
\label{related}
\subsection{Sound Event Localization and Detection}
There are some traditional sound source localization algorithms such as multiple signal classification (MUSIC) and the generalized cross-correlation phase transform (GCC-PHAT). With the development of deep learning techniques, an increasing number of learning-based methods have been developed. An overview of the parametric-based methods and the learning-based methods can be found in \cite{zhao2023audio}.
In \cite{adavanne2018sound}, SELDnet was proposed to predict the sound event class and the related positions recurrently. In \cite{adavanne2018sound}, an activity-coupled Cartesian DOA (ACCDOA) representation is proposed to combine the training objectives of class prediction and location prediction. Multi-ACCDOA proposed in \cite{shimada2022multi} extends ACCDOA to represent multiple sound events. In \cite{wang2023four}, a four-stage data augmentation method including audio channel swapping, multi-channel simulation, time-domain mixing and time-frequency masking is proposed to mitigate the data scarcity problem. In \cite{nguyen2022salsa}, Spatial cue-augmented
log-spectrogram (SALSA) is proposed to integrate time-frequency features for sound event detection, and integrate magnitude or phase difference for sound event localization. A light-weighted version of SALSA is proposed in \cite{nguyen2022salsalite}. In \cite{cao2020event}, the event-independent network employs two branches to predict the sound events and locations, respectively, and uses the permutation invariant training method to find the most possible combination. Transformer is used for SELD in \cite{schymura2021pilot}, which predicts the means and covariances of the sound event localizations via self-attention modules. In DCASE 2023 track b, $360 \degree$ videos are added to serve as complementary information. Visual bounding boxes are encoded as stacked Gaussian vectors in the baseline systems. In the state-of-the-art system\cite{wang2022nerc}, speech separation is employed to assist the SELD task and a video pixel-swapping data augmentation technique is proposed. The object detection module is used to refine the localization results by aligning the DOA lines with bounding boxes. In \cite{shimada2023zero}, zero-shot and few-shot SELD are tackled to predict the unseen sound event classes. 

\subsection{Multimodal Representation Learning}
Large pretrained models learn multi-modal representations and show strong one-shot or zero-shot capabilities on the downstream tasks.
CLIP \cite{radford2021learning} learns the visual representations under textual supervision using contrastive learning and exhibits superior zero-shot performance on downstream tasks such as classification and detection. Following the training paradigm of CLIP, GLIP \cite{li2022grounded} deals with language-aware object detection in a contrastive learning way. In addition, CLIP has been extended to other tasks such as video retrieval \cite{luo2022clip4clip} and action recognition \cite{wang2021actionclip}. AudioCLIP \cite{guzhov2022audioclip} extends CLIP to the audio domain and shows competitive performance on the environmental sound classification. In addition to the CLIP-based model, there are other models for multimodal representation learning. In \cite{kim2021vilt}, ViLT learns visual-textual representation with lightweight modality embedding and emphasizes modality interactions. In \cite{wang2022image}, BEiT-3 jointly trains image, texts and image-text pairs with masked modeling and employs different experts for different tasks. 
\section{Text Queried SED}
\label{define}
In this section, we define the proposed novel task. Given a text prompt describing the sound event like \textit{`A vehicle is accelerating'} and a mixed audio $\mathbf{a} \in \mathbb{R}^{C \times T}$, where $C$ is the number of channels and $T$ is the audio length, the objective is to predict the DOA $d$ of the target sound event. For static targets, we predict a single DOA value. For moving targets, we predict the trajectory of the sound event $\mathbf{d} \in \mathbb{R}^{T}$. 

\section{Model}
\label{m}
\begin{figure}[tbp]
    \centering
    \includegraphics[width=1\columnwidth]{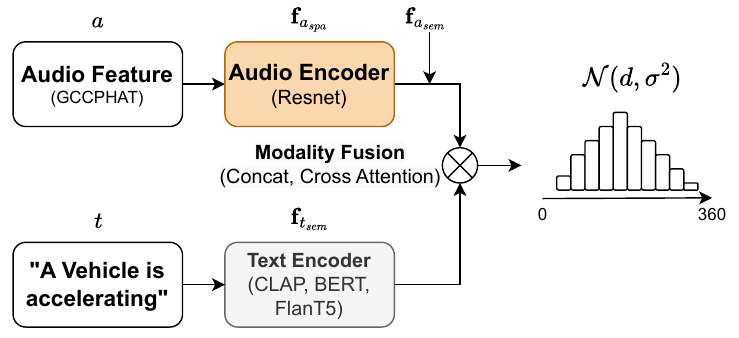}
    \caption{The proposed model consists of three parts, an audio encoder, a text encoder and the multimodal fusion module.}
    \label{model}
\end{figure}

The overall workflow of the proposed model is shown in Fig. \ref{model}, which contains three main parts, audio feature encoding, textual feature encoding and multimodal fusion. 

\subsection{Audio Feature Encoding}
We employ GCC-PHAT $\mathbf{g}(t, \tau)$ as the frame-level audio feature, which contains the spatial information and the time lag between microphone pairs can be inferred by the response map. It calculates the generalized cross-correlation of the paired audio $(m, n)$ and is normalized by the magnitude, only retaining the phase information.
\begin{equation}
    \mathbf{g}(t, \tau)=\int_{-\infty}^{+\infty} \frac{\Phi_m(t, f) \Phi_n^*(t, f)}{\left|\Phi_m(t, f)\right|\left|\Phi_n^*(t, f)\right|} e^{j 2 \pi f \tau} d f
\end{equation}
where $\Phi$ denotes the short-term Fourier transform, $\tau$ is the time lag, $*$ denotes the complex conjugate and $f$ is the frequency. Here, $\mathbf{g}(t, \tau) \in \mathbb{R}^{T \times F}$, where $T$ is the number of time frames and $F$ is the number of time lag intervals. A multi-channel feature with dimension $\mathbb{R}^{L \times T \times F}$ can be formed by stacking $\mathbf{g}$ from different microphone pairs, where $L$ is the number of microphone pairs.

To obtain the clip-level spatial feature $\mathbf{f}_{a_{spa}}$, we use ResNet18 \cite{he2016deep} to encode the frame-level feature $\mathbf{g}$ and we extract features before the classification layer $\mathbf{f}_{a_{spa}} \in \mathbb{R}^{512}$ as follows,
\begin{equation}
    \mathbf{f}_{a_{spa}} = \mathcal{F}_{ResNet}(\mathbf{g})
\end{equation}

In addition to the spatial feature, we use CLAP \cite{wu2023large} audio encoder $\mathcal{F}_{CLAPaudio}$ to obtain the clip-level semantic feature $\mathbf{f}_{a_{sem}} \in \mathbb{R}^{512}$ as follows
\begin{equation}
    \mathbf{f}_{a_{sem}} = \mathcal{F}_{CLAPaudio}(\mathbf{a})
\end{equation}

\subsection{Text Feature Encoding}
For clip-level textual feature, similar to the semantic audio feature, we employ the CLAP text encoder for extracting sound event-related information $\mathbf{f}_{t_{sem}} \in \mathbb{R}^{512}$:
\begin{equation}
    \mathbf{f}_{t_{sem}} = \mathcal{F}_{CLAPtext}(\mathbf{a})
\end{equation}

\subsection{Multimodal Fusion}
For static sound events, we make clip-level predictions and use simple concatenations to fuse the audio and text information, which proves to be effective in target speaker extraction tasks \cite{hao2023typing}, as follows  
\begin{equation}
    \mathbf{f} = [\mathbf{f}_{a_{spa}}; \mathbf{f}_{a_{sem}}; \mathbf{f}_{t_{sem}}]
\end{equation}

For moving sound events, we make frame-level predictions. We employ the Transformer \cite{vaswani2017attention} for fusion and use the frame-level audio feature $\mathbf{g}$. To ensure the time consistency between the input feature and the output, the audio feature $\mathbf{g}$ is used as the query ($\mathbf{Q}$) and the text feature is used as the value ($\mathbf{V}$) in the cross attention $CA(\mathbf{Q}, \mathbf{K}, \mathbf{V})$ module, i.e.
\begin{equation}
    \mathbf{f} = CA(\mathbf{g},\mathbf{f}_{t_{sem}}, \mathbf{f}_{t_{sem}})
\end{equation}

\subsection{Training objectives}

The DOA estimate can be represented as a one-hot vector with $\mathbb{R}^{360}$ and the model can be trained using cross entropy loss. However, cross entropy is generally used for classification tasks, treating each degree as an independent class. Here, we employ the earth mover's distance (EMD) loss \cite{yu2021metricnet}, which can model the relationship of different degrees. First, we encode the target DOA $d$ to a Gaussian distribution as the ground truth.
\begin{equation}
    y_{i} \sim \mathcal{N}(d, \sigma^2)
\end{equation}
where $1 \leq i \leq 360$, $d$ is the mean and $\sigma^2$ is a predefined covariance. 

Let $\mathbf{o} \in \mathbb{R}^{360}$ denote the output of the text queried SED system. After normalizing $\mathbf{o}$ by \textit{softmax}, the EMD loss is calculated as the accumulated difference between the encoded ground truth and $\mathbf{o}$, which measures the similarities between two discrete distributions, as follows 
\begin{equation}
    loss_{EMD} = \sum _{i = 1}^{360} 
    \left|softmax({o}_{i}) - {y}_{i}\right|
\end{equation}
where $o_{i}$ is the $i$-th element of $\mathbf{o}$.   
\section{Implementation Details}
\label{implement}
For calculating STFT, $n\_fft$ is set to 1024 with 640 hop size. The number of time lags in GCC-PHAT is set to 96. For extraction of audio and text semantic features, we use CLAP encoder in \cite{wu2023large} to obtain $\mathbf{f}_{a_{sem}}$ and $\mathbf{f}_{t_{sem}} \in \mathbb{R}^{512}$. For multimodal fusion, we use 4 encoders and 4 decoders in the transformer with 256 hidden dimensions and 1024 feedforward dimensions. For model training, the batch size is set to 64 with an initial learning rate of 5e-4. The learning rate decays by 0.5 after 20 epochs. The training process is monitored by the early stop mechanism with the patience of 10. The covariance in the EMD loss function is set to 5.

We use mean average error (MSE) to measure the model performance, which is calculated as follows. 
\begin{equation}
    MSE = \frac{1}{T}\sum_{t = 0} ^{T} \left | \argmax_{i}(\mathbf{o}_{i, t}) - d_{t} \right |
\end{equation}
where $T$ is the number of time steps. The predicted azimuth is derived by $\argmax$ for the predicted distribution. As the azimuth is continuous between 360$\degree$ and 1$\degree$, we use the remainder of 360$\degree$ if the error is larger than 180$\degree$. 

\begin{figure}[tbp]
    \centering
    \includegraphics[width=0.9\columnwidth]{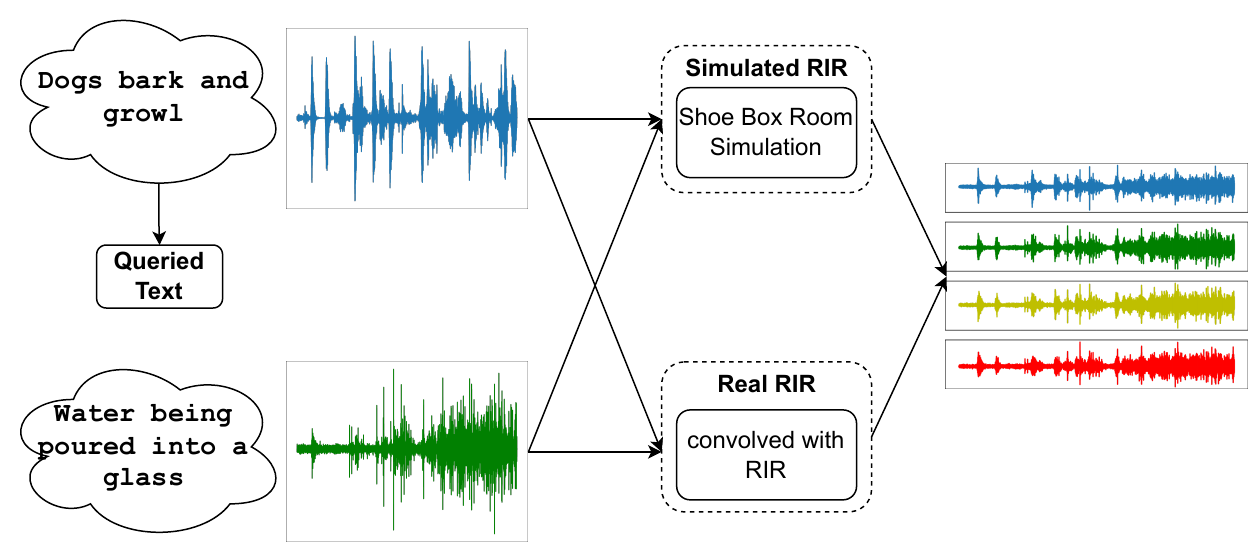}
    \caption{The simulation process using simulated RIR (the upper branch) and real RIR (the bottom branch).}
    \label{data}
\end{figure}

\section{Experiment Analysis}
\label{experiment}
We evaluate the proposed methods in two ways. We perform evaluations for the baseline methods on the simulated dataset created by simulated RIR, where the sound sources are static.  The dataset is simulated on the fly from AudioCaps \cite{kim2019audiocaps}. In addition, the baseline methods are evaluated on the dataset created by real RIR, where the sound sources are either static or moving. Each audio clip is truncated or padded to 10 seconds. We follow the original dataset split of AudioCaps to generate the training, evaluation and test set. For both cases, we randomly choose two audio clips with their corresponding captions and randomly choose one caption for the query. The simulation process in both cases is shown in Fig. \ref{data}.

\subsection{Evaluation on Simulated RIR}

The simulated dataset is created by Pyroomacoustics \cite{scheibler2018pyroomacoustics} and the RIR is simulated by the image source model. The length, width and height of the room range from 5 meter to 15 meter. RT60 is randomly chosen in the range from 0.5 second to 1 second. The shape of the microphone array is a square with four microphones with a height from 1 meter to 1.2 meter. The inter-distance of each microphone ranges from 0.1 meter to 0.13 meter. 

\subsection{Evaluation on Real RIR}
To evaluate the model performance on real scenarios, we also simulate the dataset using real RIR, which is created by the DCASE data generator \cite{politis2021dataset} \footnote{https://github.com/danielkrause/DCASE2022-data-generator}. It is widely used in many SELD systems \cite{wang2022nerc} for pretraining. The RIRs are collected in 10 more rooms in the Tampere University and different rooms have different reverberation conditions. The RIRs recorded in rooms 1, 2, 3, 4, 5, 6 and 10 are used for generating the training data. The RIRs recorded in rooms 8 and 9 are used for evaluation and test data generation, respectively. The data generator will produce the format of FOA and MIC multi-channel audio and we only adopt the FOA format, which is 4-channel spatial audio converted from 32-channel Eigenmike format \cite{politis2021dataset}. We generate 7224, 903 and 903 audio clips for training, evaluation and testing, respectively.

\begin{table}[tbp]
\caption{Experimental results on the dataset created by simulated RIR, where sound events are static. `dir.' denotes the directional source and `add.' denotes the additive source. `CA' means cross attention.}
\centering
\setlength{\tabcolsep}{1mm}{

\begin{tabular}{lllll}
\toprule[1pt]
\textbf{Text Feature} & \textbf{Audio Feature}                                            & \textbf{Fusion}          & \textbf{Datasets}                                                                              & \textbf{MAE} ($\degree$)   \\ \hline\specialrule{0em}{1pt}{1pt}
CLAP         & GCC-PHAT                                                 & Concat          & \begin{tabular}[c]{@{}l@{}}One dir.,\\ One add.\end{tabular} & 1.34  \\ \hline\specialrule{0em}{1pt}{1pt}
CLAP         & GCC-PHAT                                                 & Concat          & Two dir.                                                               & 32.15 \\ \hline\specialrule{0em}{1pt}{1pt}
CLAP         & \begin{tabular}[c]{@{}l@{}}GCC-PHAT,\\ CLAP\end{tabular} & Concat          & Two dir.                                                               & 43.3  \\ \hline\specialrule{0em}{1pt}{1pt}
FlanT5       & GCC-PHAT                                                 & CA & Two dir.                                                              & 37.18 \\ \hline\specialrule{0em}{1pt}{1pt}
BERT         & GCC-PHAT                                                 & CA & Two dir.                                                              & 33.61 \\ \bottomrule[1pt]
\end{tabular}
}
\label{t1}
\end{table}

\begin{table}[tbp]
\caption{Experimental results on the dataset created by real RIR, where sound events are either static or moving. }
\centering
\setlength{\tabcolsep}{1mm}{
\begin{tabular}{llll}
\toprule[1pt]
\textbf{Text Feature} & \textbf{Audio Feature}                                   & \textbf{Fusion} & \textbf{MAE} ($\degree$) \\ \hline\specialrule{0em}{1pt}{1pt}
CLAP                  & GCC-PHAT                                                 & Concat          & 45.58        \\ \hline\specialrule{0em}{1pt}{1pt}
CLAP                  & \begin{tabular}[c]{@{}l@{}}GCC-PHAT,\\ CLAP\end{tabular} & Concat          & 47.92        \\ \hline\specialrule{0em}{1pt}{1pt}
FlanT5                & GCC-PHAT                                                 & CA              & 48.71        \\ \hline\specialrule{0em}{1pt}{1pt}
BERT                  & GCC-PHAT                                                 & CA              & 49.18        \\ \bottomrule[1pt]
\end{tabular}}
\label{t2}
\end{table}


\subsection{Experimental Results}
The MAE results on the simulated dataset using simulated RIR are shown in TABLE \ref{t1}. As there are no previous work on text-queried target sound source localization, we compare different combinations of audio feature, text feature and fusion methods. We compare clip-level fusion and frame-level fusion. Firstly, we evaluate the clip-level fusion model using CLAP + GCC-PHAT with concatenation on a simple scenario where there is only one directional sound source, and the other sound source is added as background noise. A fully connected layer is used as a classifier to obtain $o$, with the concatenated features as the input. It can be seen that the model can accurately locate the sound source with 1.34$\degree$ MAE error. When there are two directional sources, it is more difficult to localize the target sound source and the MAE error is larger. CLAP + GCC-PHAT with concatenation performs best and adding CLAP audio embedding does not improve the localization performance. In addition, we use frame-level cross attention fusion with FlanT5 \cite{chung2022scaling} and BERT \cite{devlin2018bert} text embedding as key. The maximium text length is set to 15 and the embedding dimensions are $15 \times 1024$ and $15 \times 768$ with corresponding masks. A [CLS] token is added before the GCC-PHAT along the time dimension. The output token corresponding to the [CLS] token is used for prediction. However, the performance is not as good as that of the clip-level fusion method.

\begin{figure}[tbp]
    \centering
    \includegraphics[width=0.9\columnwidth]{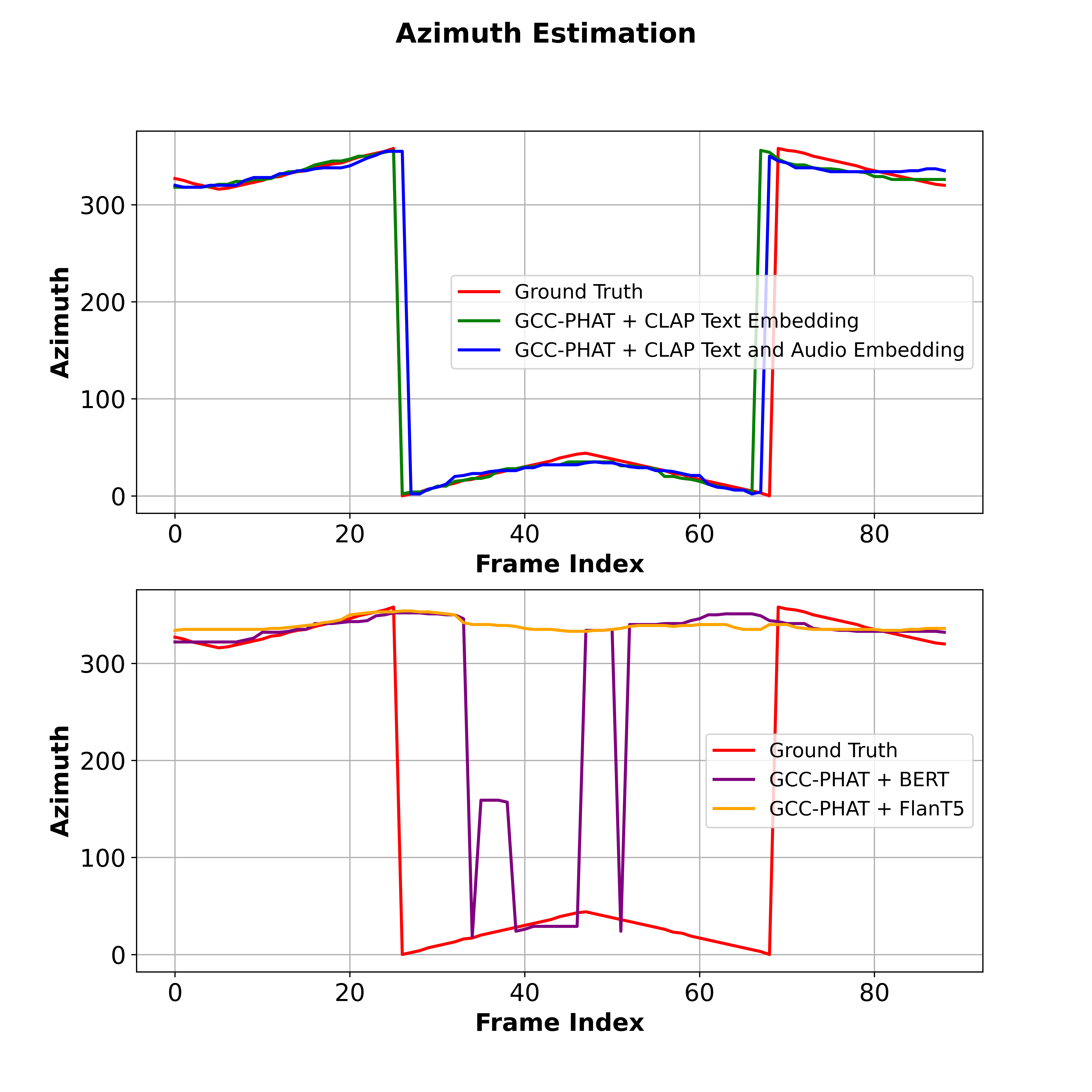}
    \caption{Visualization results of the clip-level fusion methods (the upper figure) and the frame-level fusion methods (the bottom figure).}
    \label{visual}
\end{figure}

Evaluation results on the dataset created by real RIR are shown in TABLE \ref{t2}. Different from the dataset created simulated RIR, the sound sources are either moving or static and the target location is predicted at frame-level. 
We also explore the fusion methods such as concatenation and cross attention. For concatenation, LSTM is employed to take the concatenated input for frame-level prediction. 
It is clear that the dataset created by real RIR with moving trajectories is more challenging and the MAE error is larger than 40$\degree$. We visualize the localization results in Fig. \ref{visual}. The data sequence is generated by fold 8 through the DCASE data generator. For the given sequence, it can be seen that the performance of the clip-level fusion method is better than that of the frame-level method. From frame 28 to frame 58, the sound source moves from 340$\degree$ to approximately 40$\degree$. The clip-level fusion based methods are able to capture the movement while the frame-level fusion based methods fail.

\section{Conclusion}
\label{con}
In this paper, we have presented a text-queried target sound source localization method and perform a benchmark study to explore the performance of the multimodal fusion baselines. Experimental results show that the concatenation of CLAP text embedding and GCC-PHAT achieves the best localization performance. As there are no existing datasets on the proposed novel task, we only evaluate the model performance on the simulated datasets.  
For future work, we will evaluate the model performance on real scenarios and improve the localization accuracy.

\section{Acknowledgement}

This research was sponsored in part by Tencent AI Lab Rhino-Bird Gift Fund and in part by University of Surrey. The work of Xinyuan Qian was sponsored in part by CCF-Tencent Rhino-Bird Open Research Fund and National Natural Science Foundation of China, Grant No. 62306029.
\bibliographystyle{IEEEtran}
\bibliography{ref}

\end{document}